\documentclass[twocolumn,showpacs,preprintnumbers,amsmath,amssymb,superscriptaddress,prb]{revtex4}

\usepackage{graphicx,here}
\usepackage{dcolumn}
\usepackage{bm}
\usepackage{enumerate}

\begin{document}

\preprint{APS/123-QED}

\title{Control of ferroelectric polarization via uniaxial pressure\\
 in the spin-lattice-coupled multiferroic CuFe$_{1-x}$Ga$_x$O$_2$}

\author{T. Nakajima}
\email{E-mail address: nakajima@nsmsmac4.ph.kagu.tus.ac.jp}
\author{S. Mitsuda}
\author{T. Nakamura}
\author{H. Ishii}
\author{T. Haku}
\author{Y. Honma}
\affiliation{Department of Physics, Faculty of Science, Tokyo University of Science, Tokyo 162-8601, Japan}%
\author{M. Kosaka}
\affiliation{Graduate School of Science and Engineering, Saitama University, Saitama, Saitama 338-8570, Japan}
\author{N. Aso}
\affiliation{Faculty of Science, University of the Ryukyus, Nishihara, Okinawa 903-0213, Japan}
\author{Y. Uwatoko}
\affiliation{Institute for Solid State Physics, University of Tokyo, Kashiwa, Chiba 277-8581, Japan}

\begin{abstract}
We have demonstrated that ferroelectric polarization in a spin-driven multiferroic CuFe$_{1-x}$Ga$_x$O$_2$ with $x=0.035$ can be controlled by the application of uniaxial pressure. %
Our neutron diffraction and in-situ ferroelectric polarization measurements have revealed that the pressure dependence of the ferroelectric polarization is explained by repopulation of three types of magnetic domains originating from the trigonal symmetry of the crystal. %
We conclude that the spin-driven anisotropic lattice distortion and the fixed relationship between the directions of the magnetic modulation wave vector and the ferroelectric polarization are the keys to this spin-mediated piezoelectric effect. 
\end{abstract}

\maketitle

Magnetic frustration induces a variety of cross-correlated phenomena in condensed matter. For example, frustrated magnets often exhibit anisotropic lattice distortion associated with magnetic order, in order to lift the degeneracy due to competing magnetic interactions\cite{SpinLattice1,PRL2004_Penc}. %
In other words, spin and lattice degrees of freedom are tightly coupled with each other in frustrated magnets. %
Another example is spin-driven ferroelectricity. Frustrated magnets often display complex noncollinear magnetic orderings, some of which break the inversion symmetry of a system. %
Recent studies on magneto-electric (ME) multiferroics have revealed that magnetic inversion symmetry breaking accounts for the ferroelectricity\cite{Kimura_nature,Katsura_PRL_2005,Kenzelmann_PRL_Spiral}. %
Based on the relationships among the spin, lattice and dielectric degrees of freedom in frustrated magnets, we have demonstrated that spin-driven ferroelectric polarization is controlled by pressure, which directly affects the lattice degree of freedom, in the spin-lattice-coupled ME multiferroic CuFe$_{1-x}$Ga$_x$O$_2$ (CFGO) with $x=0.035$. %

The delafossite compound CuFeO$_2$ (CFO) is a triangular lattice antiferromaget, which is a typical example of frustrated spin system. %
Because of antiferromagnetic interactions between the magnetic Fe$^{3+}$ ions forming the triangular lattice, this system has strong magnetic frustration and it therefore exhibits a variety of unconventional magnetic phase transitions\cite{Mitsuda_1991,Mitsuda_2000}. %
Despite the Heisenberg spin character expected from the electronic state of the magnetic Fe$^{3+}$ ions ($S=\frac{5}{2}, L=0$), CFO exhibits an Ising-like collinear four-sublattice (4SL) magnetic ground state. %
A small amount of nonmagnetic substitution or the application of a magnetic field turns the 4SL phase to a noncollinear incommensurate magnetic phase, in which spin-driven ferroelectricity has been discovered\cite{Kimura_CuFeO2,Seki_PRB_2007,Ga-induce}. %
Hereafter, we refer to this ferroelectric phase as the `ferroelectric incommensurate magnetic (FE-ICM) phase'. %
The magnetic structure of the FE-ICM phase has been determined to be an elliptic screw-type magnetic structure, whose magnetic modulation wave vector is $(q,q,\frac{3}{2})$ with $q=0.203$\cite{SpinNoncollinearlity,CompHelicity}. %
Spin-driven ferroelectric polarization emerges along the screw axis, which is parallel to the [110] direction of the crystal. %
The polarity of the ferroelectric polarization is coupled with the chirality of the magnetic structure, i.e., the left-handed (LH) and right-handed (RH) helical arrangements of the spins\cite{CFAO_Helicity,CompHelicity}. %

\begin{figure}[t]
\begin{center}
	\includegraphics[clip,keepaspectratio,width=7.5cm]{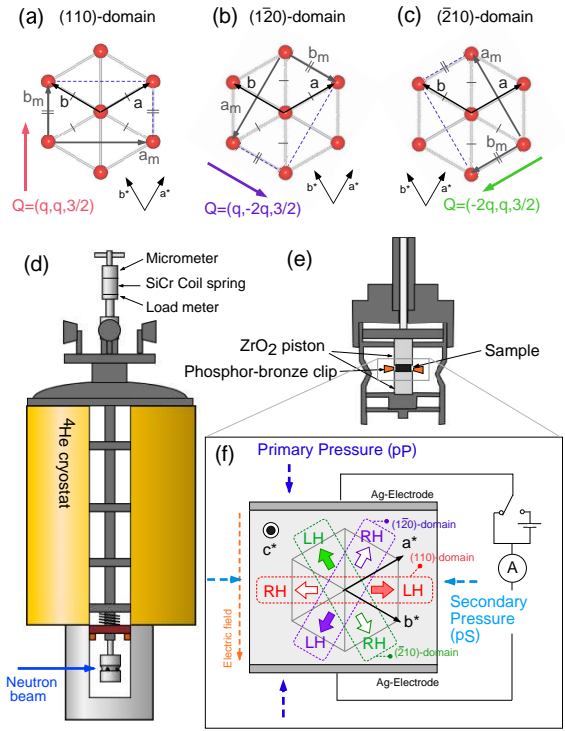}
	\caption{(Color Online) [(a)-(c)] Relationships between the hexagonal and the monoclinic bases in the three types of domains. %
	(d) Schematic drawing of the uniaxial pressure cell inserted in the cryostat. %
	(e) Magnification of the sample space of the pressure cell. %
	(f) Relationships among the six possible directions of the ferroelectric polarization, applied electric field and two directions of uniaxial pressure. Open and filled arrows denote the directions of the ferroelectric polarization induced by the RH- and the LH-helical magnetic domains, respectively.}
	\label{domains}
\end{center}
\end{figure}

CFO also exhibits spin-driven crystal structural transitions. Recent synchrotron radiation x-ray diffraction studies have revealed that CFO exhibits monoclinic lattice distortions in several magnetically ordered phases including the FE-ICM phase\cite{Terada_CuFeO2_Xray,Ye_CuFeO2,Terada_14.5T,CFAO_Xray}, while CFO has a trigonal crystal structure in the paramagnetic (PM) phase. %
This spin-driven structural transition produces three types of monoclinic crystal domains because the original trigonal crystal structure has a threefold rotational symmetry about the $c$ axis. %
The relationships between the hexagonal and the monoclinic bases in each domain are shown in Figs. \ref{domains}(a)-\ref{domains}(c). %
Note that in this letter, we have mainly employed the hexagonal basis to describe the three types of domains. We have added subscripts ``m'' when referring to the monoclinic bases. %
In each monoclinic domain, the $b_{\rm m}$ axis elongates and, in contrast, the $a_{\rm m}$ axis contracts, compared to those in the PM phase. In addition, the $c^*$-plane-projection of the magnetic modulation wave vector, which is parallel to the screw axis in the FE-ICM phase, is fixed to be parallel to the $b_{\rm m}$ axis. %
Hereafter, we refer to the three types of coupled magnetic and crystal domains as $(110)$-, $(1\bar{2}0)$- and $(\bar{2}10)$-domains, as shown in Figs. \ref{domains}(a)-\ref{domains}(c). %
Taking account of the strong spin-lattice coupling in this system, we have recently demonstrated that in CFO, the volume fractions of the three magnetic domains can be controlled by uniaxial pressure applied perpendicular to the $c$ axis\cite{SingleDomainSW}. %
We thus expect that in the FE-ICM phase, application of uniaxial pressure results in distinct changes in the ferroelectric polarization reflecting the changes in the volume fractions of the three magnetic domains. %
In the present study, we have performed neutron diffraction and in-situ ferroelectric polarization measurements under applied uniaxial pressure, using a CFGO ($x=0.035$) sample, which exhibits the FE-ICM phase below $\sim$8 K in zero magnetic field.

A single crystal of CFGO($x=0.035$) of nominal composition was prepared by the floating zone method\cite{Zhao_FZ} and %
was cut to dimensions of $\sim2.9\times2.1\times4.0$ mm$^3$ with the $[1\bar{1}0]$ direction parallel to the shortest dimension. %
We have developed a uniaxial pressure cell loaded into a pumped $^4$He cryostat, as illustrated in Fig. \ref{domains}(d). %
The uniaxial pressure along the vertical direction is controlled by a SiCr coil spring and a micrometer attached on top of the cryostat and is monitored by a load meter. %
The sample was mounted in the pressure cell with the $(H,H,L)$ scattering plane orientated so that %
the vertical uniaxial pressure was applied to the $[1\bar{1}0]$ surfaces of the sample. %
Uniaxial pressure was also applied to the [110] surfaces by a small phosphor-bronze clip. %
Hereafter, we refer to the vertical pressure applied by the spring and the horizontal pressure applied by the clip as `primary pressure' ($p_P$) and `secondary pressure' ($p_S$), respectively. %
In this experimental setup, we can control the magnitude of $p_P$ even at low temperatures. On the other hand, we cannot control $p_S$, %
which was kept applied throughout the present experiment. %
The magnitude of $p_S$ is assumed to be relatively small. %
Silver paste electrodes were applied to the $[1\bar{1}0]$ surfaces of the sample. %
The ferroelectric polarization along the $[1\bar{1}0]$ direction ($P$) was deduced by integrating the pyroelectric or piezoelectric current measured by an electrometer. %
Neutron diffraction measurements under an applied uniaxial pressure were carried out with a triple axis spectrometer HQR installed in JRR-3 of the Japan Atomic Energy Agency in Tokai, Japan. %
The wavelength of the incident neutron beam was 2.44 \AA, and the collimation was open-40'-40'-40'. %

\begin{figure}[t]
\begin{center}
	\includegraphics[clip,keepaspectratio,width=7.2cm]{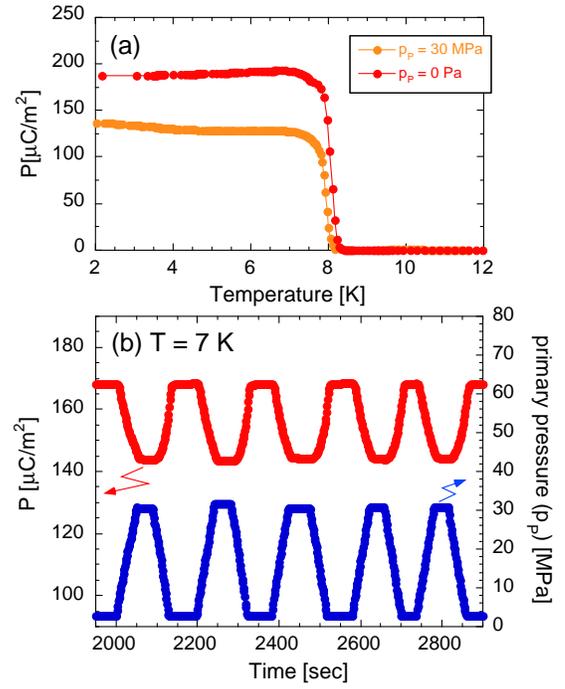}
	\caption{(Color Online) (a) Temperature variations of $P$ measured on heating at $p_P = $ 0 and 30 MPa. %
	(b) Time dependences of $P$ and $p_P$ at 7 K. }
	\label{pyro}
\end{center}
\end{figure}

Figure \ref{pyro}(a) shows the temperature variations of $P$ measured on heating at $p_P =$ 0 and 30 MPa. %
We have found that the application of only 30 MPa of $p_P$ significantly reduces the value of $P$. %
At 7 K, just below the ferroelectric transition temperature, we have measured the $p_P$-dependence of $P$ in zero electric field\cite{com1}. %
Before this measurement, the sample was cooled with an applied electric field of 116 kV/m from 10 K to 2 K, and then was heated to 7 K in zero electric field. %
In Fig. \ref{pyro}(b), we show the values of $P$ and $p_P$ as functions of time\cite{com3}, revealing that the time dependence of $P$ is synchronized with that of $p_P$: %
$P$ gradually decreases and increases with increasing and decreasing $p_P$, respectively.%

\begin{figure}[t]
\begin{center}
	\includegraphics[clip,keepaspectratio,width=7.5cm]{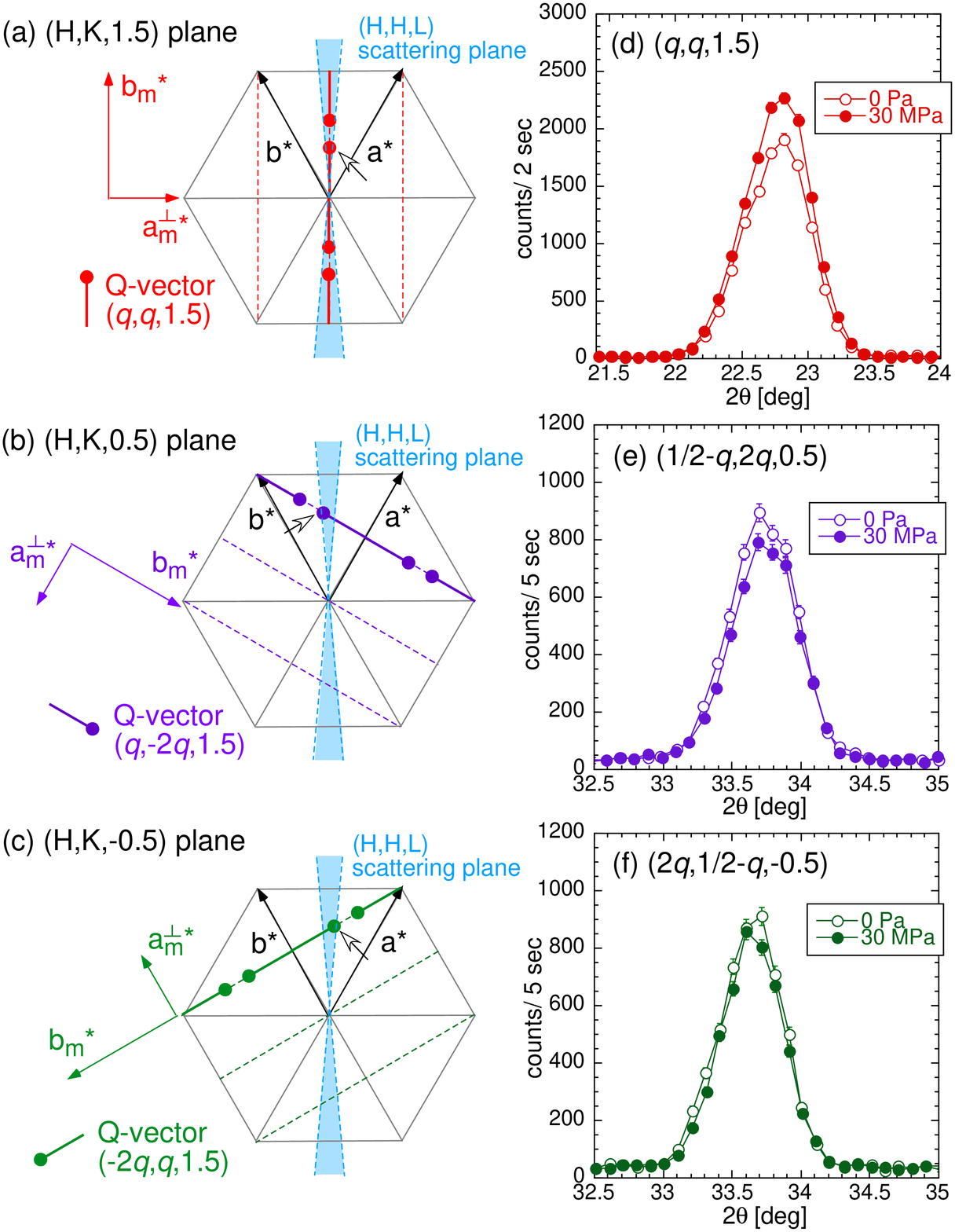}
	\caption{(Color Online) [(a)-(c)] The locations of the magnetic Bragg reflections belonging to the (a) (110)-, (b) $(1\bar{2}0)$- and (c) $(\bar{2}10)$-domains, in the reciprocal lattice space. $a_{\rm m}^{\perp*}$ denotes the $c^*$-plane-projection of $a_{\rm m}^*$. %
	Open arrows denote the positions of the magnetic reflections corresponding to (d)-(f). %
	Although the magnetic reflections belonging to the $(1\bar{2}0)$- and $(\bar{2}10)$-domains do not appear on the $(H,H,L)$ plane in principle, we can access the two reflections belonging to these domains by tilting the cryostat by $\pm \sim 4^{\circ}$. %
	[(d)-(f)] Comparisons between the magnetic diffraction profiles measured at $p_P= 0$ and $30$ MPa at 7 K }
	\label{neutron}
\end{center}
\end{figure}

Subsequent to the ferroelectric polarization measurement, we have performed neutron diffraction measurements. %
We have measured three magnetic Bragg reflections at ($q,q,\frac{3}{2}$), $(\frac{1}{2}-q,2q,\frac{1}{2})$ and $(2q,\frac{1}{2}-q,-\frac{1}{2})$, which belong to the (110)-, $(1\bar{2}0)$- and $(\bar{2}10)$-domains, respectively. %
In Figs. \ref{neutron}(d)-\ref{neutron}(f), we show the comparisons between the magnetic diffraction profiles measured at $p_P= 0$ and $30$ MPa. %
We have found that the intensity of the ($q,q,\frac{3}{2}$) reflection is enhanced by the application of $p_P$. In contrast, those of the $(\frac{1}{2}-q,2q,\frac{1}{2})$ and $(2q,\frac{1}{2}-q,-\frac{1}{2})$ reflections are reduced. %
This suggests that the volume fraction of the (110)-domain increases and those of the $(1\bar{2}0)$- and $(\bar{2}10)$-domains decrease with increasing $p_P$. %
Since the ferroelectric polarization vector in the (110)-domain is parallel to the electrodes, %
the observed value of $P$ is proportional to the sum of the volume fractions of the $(1\bar{2}0)$- and $(\bar{2}10)$-domains. Therefore, the $p_P$ dependences of the intensities of the magnetic Bragg reflections are consistent with the results of the ferroelectric polarization measurements. %

\begin{figure}[t]
\begin{center}
	\includegraphics[clip,keepaspectratio,width=7.5cm]{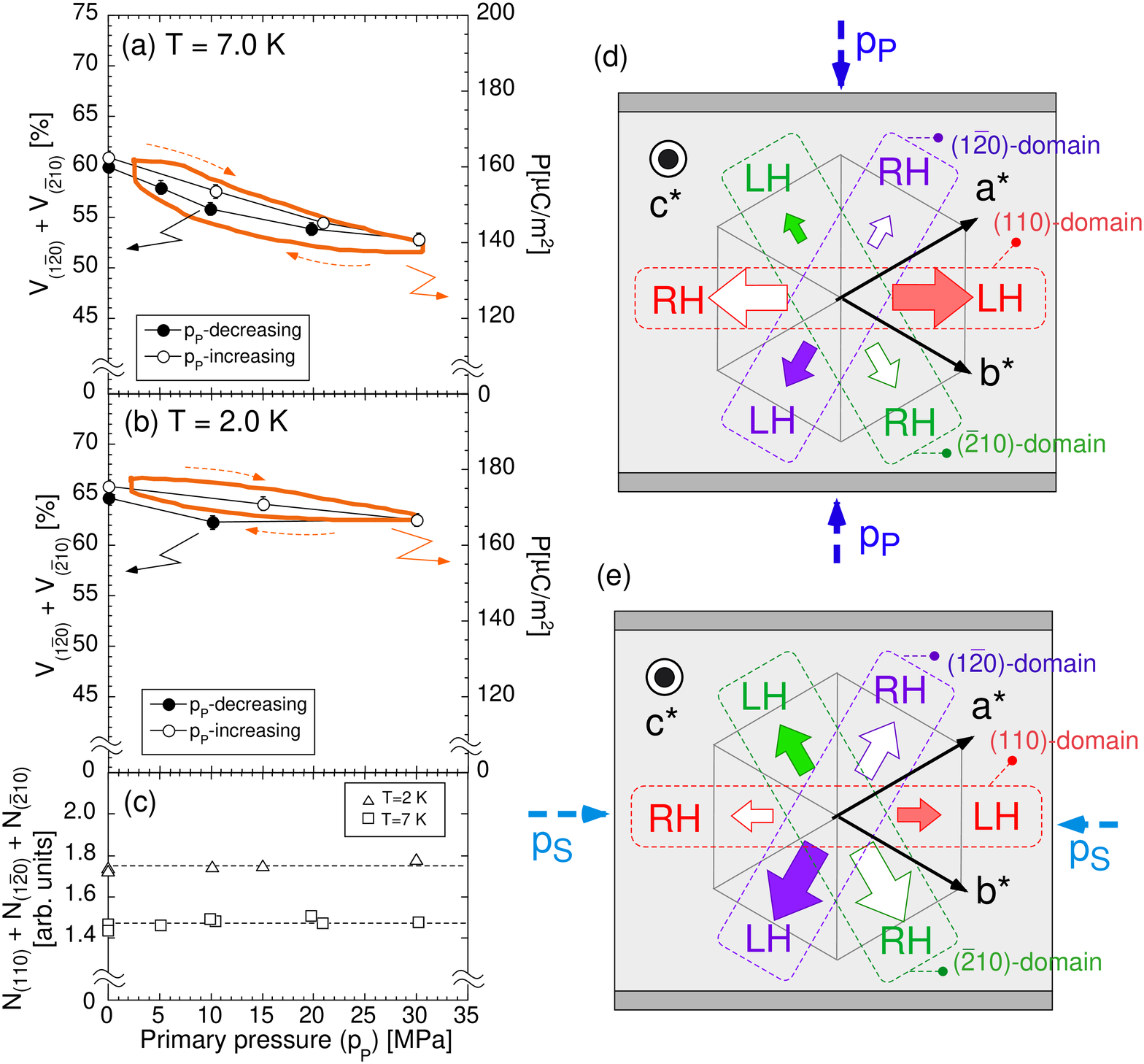}
	\caption{(Color Online) [(a),(b)] Comparison between the $p_P$ dependences of $P$ and $V_{(1\bar{2}0)} + V_{(\bar{2}10)}$ at (a) 7 K and (b) 2 K. %
	The ratio of the scales of the vertical axes ($P$ and $V_{(1\bar{2}0)} + V_{(\bar{2}10)}$) is the same in (a) and (b). %
	(c) $p_P$ dependences of $N_{(110)} + N_{(1\bar{2}0)} + N_{(\bar{2}10)}$ at 2 K and 7 K. %
	[(d),(e)] Schematic drawings of the volume fractions of the coupled magnetic and ferroelectric domains (d) when the effect of $p_P$ is dominant and (e) when the effect of $p_S$ is dominant. %
	Sizes of the open and filled arrows qualitatively show the volume fractions of the RH- and the LH-helical magnetic domains, respectively. }%
	\label{neutron_vs_pyro}
\end{center}
\end{figure}

Hereafter, we refer to the volume fractions of the (110)-, $(1\bar{2}0)$- and $(\bar{2}10)$-domains as $V_{(110)}$, $V_{(1\bar{2}0)}$ and $V_{(\bar{2}10)}$, respectively. The sum of the volume fractions relevant to $P$ is described by
\begin{eqnarray}
P\propto V_{(1\bar{2}0)} + V_{(\bar{2}10)} = \frac{N_{(1\bar{2}0)} + N_{(\bar{2}10)}}{N_{(110)} + N_{(1\bar{2}0)} + N_{(\bar{2}10)}},
\end{eqnarray}
where $N_\alpha$ [$\alpha = (110), (1\bar{2}0)$ or $(\bar{2}10)$] is the integrated intensity of the magnetic Bragg reflection belonging to the $\alpha$-domain divided by the Lorentz factor and the square of the calculated magnetic structure factor for the magnetic reflection. %
It should be noted that the sum of $N_{(110)}$, $N_{(1\bar{2}0)}$ and $N_{(\bar{2}10)}$ is independent of $p_P$, %
as shown in Fig. \ref{neutron_vs_pyro}(c). %
This suggests that the application of 30 MPa of $p_P$ does not affect the magnetic structure in the FE-ICM phase, and therefore the changes in the intensities of the magnetic Bragg reflections can be ascribed only to the changes in the volume fractions of the three domains. %
Figures \ref{neutron_vs_pyro}(a) and \ref{neutron_vs_pyro}(b) show the values of $P$ and $V_{(1\bar{2}0)} + V_{(\bar{2}10)}$ at 7 K and 2 K as functions of $p_P$.\cite{com2} %
We have found that the $p_P$-variations of $P$ agree with those of $V_{(1\bar{2}0)} + V_{(\bar{2}10)}$ at both temperatures. %
This indicates that the changes in $P$ are explained by the pressure-induced repopulation of the three magnetic domains. %
Specifically, %
the uniaxial pressure along the $[1\bar{1}0]$ direction favors the domains whose $a_{\rm m}$ axis lies along the direction of the pressure, and suppresses the other domains, because of the anisotropic lattice distortion in each domain. %
Hence, $V_{(1\bar{2}0)}$ and $V_{(\bar{2}10)}$ decrease with increasing $p_P$, as illustrated in Fig. \ref{neutron_vs_pyro}(d). %
When $p_P$ is removed and the effect of $p_S$ is dominant, $p_S$ suppresses the magnetic domains whose $b_{\rm m}$ axis are parallel to $p_S$ and enhances the other domains, and therefore $V_{(1\bar{2}0)}$ and $V_{(\bar{2}10)}$ increase with decreasing $p_P$, as illustrated in Fig. \ref{neutron_vs_pyro}(e). %
It should be noted that $V_{(1\bar{2}0)}$ and $V_{(\bar{2}10)}$ might not be retrieved without the application of $p_S$. %

We now discuss the microscopic picture of the domain structures in this system. %
The fact that the $p_P$-variations of $P$ are proportional to those of $V_{(1\bar{2}0)} + V_{(\bar{2}10)}$ %
indicates that the asymmetry of the LH- and the RH-helical magnetic orderings in each of the $(110), (1\bar{2}0)$ or $(\bar{2}10)$-domains remain unchanged even after the volume fractions of the three domains change. %
This implies that %
the domain walls separating the three types of crystal domains gradually move with increasing or decreasing pressure, keeping the asymmetry of the LH- and the RH-helical magnetic orderings in each of the domains. %
In other words, sudden flops or reconstruction of the crystal domains, in which the magnetic chirality might be no longer conserved, hardly occur in the present experiments. %
The present results also suggest that mobility of the domain walls are reduced at low temperatures. %

In conclusion, we have demonstrated uniaxial pressure control of ferroelectric polarization in a frustrated magnet CuFe$_{1-x}$Ga$_x$O$_2$ with $x=0.035$, in which the spin, lattice and  dielectric degrees of freedom are tightly coupled with each other. %
The present neutron diffraction and in-situ ferroelectric polarization measurements have revealed that the application of uniaxial pressure along the $[1\bar{1}0]$ direction results in repopulation of the three types of coupled crystal and magnetic domains owing to the spin-driven anisotropic lattice distortions in each domain. %
Consequently, the ferroelectric polarization along the $[1\bar{1}0]$ direction changes with the applied pressure, reflecting the domain structure in the sample. %
Although the present work has focused only on the pressure dependence of the domain structure, we expect that spin-frustrated magnets have the potential to show a variety of pressure-induced cross correlated phenomena including this kind of spin-mediated piezoelectric effect.

\section*{ACKNOWLEDGMENTS}
The neutron diffraction measurements at JRR-3 were carried out as part of proposal 10695B and were partly supported by %
ISSP of the University of Tokyo. 

\bibliography{main}

\begin{thebibliography}{22}
\bibitem[{com({\natexlab{a}})}]{com1}
\bibinfo{note}{After the piezoelectric and the neutron diffraction measurements
  shown in Figs. \ref{pyro}(b), \ref{neutron_vs_pyro}(a) and
  \ref{neutron_vs_pyro}(b), we have performed pyroelectric measurements with
  increasing temperature beyond $T_c$ to determine the absolute values of $P$
  in Figs. \ref{pyro}(b), \ref{neutron_vs_pyro}(a) and
  \ref{neutron_vs_pyro}(b).}
\bibitem[{com({\natexlab{b}})}]{com3}
\bibinfo{note}{The measurements of Figs. \ref{pyro}(b) and \ref{neutron} were
  performed after several pressure cycles at 7 K. After the first pressure
  cycle, the maximum value of the ferroelectric polarization (at zero pressure)
  decreased from $\sim185$ $\mu$C/m$^2$ to $\sim170$ $\mu$C/m$^2$, but it
  didn't change in the 2nd cycle and later.}
\bibitem[{com({\natexlab{c}})}]{com2}
\bibinfo{note}{The $p_P$-variations of the volume fractions shown in Figs.
  \ref{neutron_vs_pyro}(a) and \ref{neutron_vs_pyro}(b) are rather small as
  compared to those in the previous study\cite{SingleDomainSW}. This is because
  the effect of the secondary pressure (not applied in the previous work)
  conflicts with that of the primary pressure. As a result, the sensitivity to
  $p_P$ is reduced.}
\end{thebibliography}

\end{document}